\documentclass[epj]{svjour}

\usepackage{amssymb}
\usepackage{amsmath}
\usepackage{graphicx}
\usepackage{subfigure}
\def\be{\begin{equation}}
\def\ee{\end{equation}}
\def\bea{\begin{eqnarray}}
\def\eea{\end{eqnarray}}

\begin{document}





\title{The antiferromagnetic/paramagnetic transition in mixed-spin  
compounds R$_{2}$BaNiO$_{5}$ }


\author{J.V. Alvarez\inst{1} \and Roser Valent{\'{\i}}\inst{2} }

\institute{Department of Physics, University of Michigan.
         Ann Arbor 48109, MI, USA. \and
       Institut f\"ur Theoretische Physik
                 J.W.G-Universit\"at Frankfurt,
                 Robert-Mayer-Str. 8
                60054 Frankfurt am Main, Germany.}
\date{}

\abstract{
We present an extensive Quantum Monte Carlo study of 
 the magnetic properties of
the mixed-spin quantum systems $R_2$BaNiO$_5$  ( $R$= magnetic rare
earth)
which show coexistence
of  3-dimensional  magnetic long-range order
with 1-di\-men\-sional  quantum gap excitations.
We discuss the validity of the performed simulations in the
critical region and show the excellent agreement  with experimental results. 
We emphasize the importance of quantum fluctuations contained in our
study
which is absent in previous  mean-field-like treatments.  } 

\maketitle
%
\section{Introduction}
\label{intro}
 The competition of different orders in mixed-species systems is
a subject of actual interest, not only in condensed matter physics
where for instance mixed-spin systems have been studied intensively
 in the last years\cite{Alvarez_04}
but also in other research areas like the physics of ultra-cold
atomic and molecular gases where Fermi-Bose mixtures are being
 investigated\cite{Lewenstein_04}.  The fascination in the mixed-species
 systems is that they have a very rich phase diagram with coexistence
 of different orders what leads to interesting competing effects. 

Here we want to concentrate on the magnetic behavior of a
class of mixed-spin quantum systems, R$_2$BaNiO$_5$ (R=magnetic rare earth)
 which
show coexistence of  3-dimensional (3D) magnetic long-range order
with 1-di\-men\-sional (1D) quantum gap excitations \cite{ZHELUDEV98}.
In a previous publication \cite{Alvarez_02} we proposed a microscopic
model in order to describe this coexistence of {\it classical}
with {\it quantum} features  in  Nd$_2$BaNiO$_5$  which we
evaluated with the Quantum Monte Carlo (QMC) method.  In the present 
article  we  extend our previous work to the region of critical
behavior of the whole family R$_2$BaNiO$_5$
and present a detailed analysis of the QMC simulations which
was not included in \cite{Alvarez_02}.

\begin{figure}
 \includegraphics[width=0.4\textwidth]{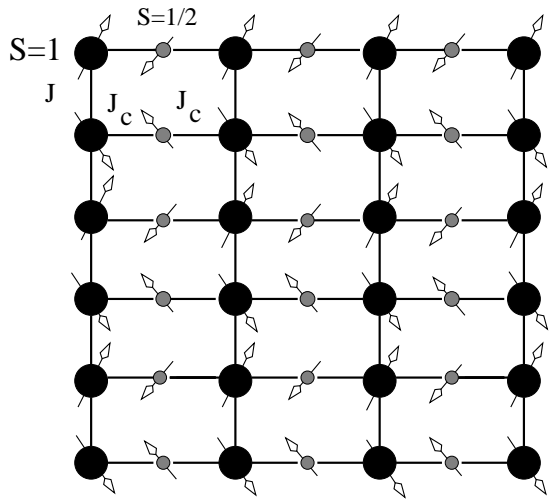}
 \caption{\label{model} Lattice picture of the model considered
 for  R$_2$BaNiO$_5$ where the
big (small) circles correspond to the $S=1$ Ni spins 
 ( $s=\frac{1}{2}$ R spins) respectively. J and J$_c$ are the Ni-Ni and the Ni-R
exchange couplings respectively.}
 \end{figure}

The R$_2$BaNiO$_5$ systems have two types of spin carriers i.e. $S=1$
Ni$^{2+}$ ions which form  antiferromagnetic chains running
along the $a$ axis of the crystal structure and $s=\frac{1}{2}$
 R$^{3+}$ ions (R= rare-earth), positioned between the chains. While 
perfect isolated  $S=1$ chains  will behave as a Haldane
 system \cite{HALDANE83}  with no long-range order even at $T=0$ and a gap
in the magnetic excitation spectrum, long-range
ordering  occurs in R$_2$BaNiO$_5$ at low temperatures  induced by
  the magnetic $s=\frac{1}{2}$ R$^{3+}$ ions.  Interestingly, 
the semi-classical 3D long-range order (i.e. spin-waves)
coexists with quantum Haldane-gap excitations in these
systems. This behavior
has been intensively studied in recent years both
experimentally   \cite{ZHELUDEV98,RAYMOND99}, as well as
 theoretically \cite{ZHELUDEV98,Alvarez_02,ZHELUDEV00}.
A first attempt to understand this coexistence of classical
and quantum orders
was done by considering a mean field approach \cite{ZHELUDEV98,ZHELUDEV00} 
 for the Ni-chains interaction which proved to
 be very meaningful to understand the basic phenomenon
of coexistence. This type of approach is nevertheless  unable
to provide a detailed des\-cription of the physics involved
and one has to invoke the use of a microscopic model which
 should include the main interactions responsible for the
system behavior.  In \cite{Alvarez_02} we considered
a spin model which describes  both the interaction within
the $S=1$ chains and  the interaction between the
 $S=1$ chains and the $s=\frac{1}{2}$ ions in between the chains
in the following way:

         \begin{eqnarray}
        H=J \sum_{ij}{\bf S}_{i, 2j}{\bf S}_{i+1, 2j}+
         J_{c}\sum_{ij} S^{z}_{i,
           2j}(s^{z}_{i, 2j-1}+s^{z}_{i, 2j+1})
      \label{Hamiltonian}
         \end{eqnarray}
with $J > 0$ and $J_c > 0$, $S$ denotes spin 1 and $s$ denotes spin
1/2. 
 The index $i$ runs along the chain direction
and $j$ in the direction perpendicular to the chains. In Fig.\ (\ref{model})
we show a schematic representation of the lattice model.  We note
that the coupling between S=1 and s=1/2 ions has been chosen
Ising-like since neutron scattering experiments\cite{Maslov_98} on these
compounds show that the excitations associated with the rare earths are
dispersionless, what indicates that the coupling between the Ni and R
sublattices must be extremely anisotropic and can therefore be approximated by an
Ising-type term.


Even very simple, minimal models are hard to evaluate
and one has to either consider some limiting assumptions
in order to be able to solve the model analytically or make
 use of numerical simulations, which may prove very accurate
 for certain models and properties. In our case, we consider
the Quantum Monte Carlo (QMC) Method.

The Hamiltonian (\ref{Hamiltonian}) has been elaborated 
in such a way that its physical properties are, by construction, 
in qualitative agreement with a wide range  
of  observations gathered in  the family of
rare earth nickelates R$_{2}$BaNiO$_{5}$. It has been shown 
\cite{Alvarez_02} that quantitative agreement to experimental results 
for Nd$_{2}$BaNiO$_{5}$ can also be achieved by  
selecting the appropriate value of  $\frac{J_c}{J}$. A first instance of this 
agreement is the ratio between the two basic energy scales involved 
in the problem: i.) The intrinsic gap $\Delta$ of the system of 
independent S=1 chains 
(when $J_{c}=0$) and ii.) the N\'eel temperature $T_{N}$ 
of the complete system when the coupling Ni-Nd is switched on.

In the well-studied Y$_{2}$BaNiO$_{5}$ \cite{YBaNiO},
the rare-earth Y$^{3+}$ is non-magnetic and no indication
of 3D long-range order has been observed, what strongly suggests
 that the system can be described by well isolated S=1 Ni$^{2+}$
chains with a Haldane gap  $\Delta \sim 127$K.  In the magnetic
  Nd$_{2}$BaNiO$_{5}$ \cite{ALONSO90,GARCIA-MATRES93,ZHELUDEV96}
the N\'eel ordering is at  $T_{N}=49$K .  Using these
 quatities we get a 
 ratio $r=\frac{\Delta}{T_{N}}=2.59$. 
Analogously, the gap value of an independent S=1 chain 
is given uniquely in terms of the
exchange constant $\Delta=0.410J$. 
For $J_{c}=0.31J$,
 we find that the ordering 
temperature is $T_{N}=0.163J$ which gives $r=2.51$ 
(a 3$\%$ difference with respect to the experimental value).
Furthermore, for the same value of the transverse coupling
$J_{c}=0.31J$,
 the staggered magnetization for the Ni sublattice
$M_{\rm Ni}(T \rightarrow 0)=0.79$ which corresponds to the value
1.6$\mu_B$ observed experimentally.
A further validation of this microscopic model was obtained
from the comparison of the calculated
staggered magnetization as a function of the temperature with
the experimental one. Actually,
the staggered magnetizations obtained from the QMC computations
for  both the  Nickel  ($M_{\rm Ni}$) and rare earth  ($M_{\rm
R}$)  sublattices turn out to be in very good quantitative 
agreement with the experimental
results for  Nd$_{2}$BaNiO$_{5}$ \cite{ZHELUDEV98,Alvarez_02} 
for $J_{c}=0.31J$. 

 Here, based on the model proposed in our previous work \cite{Alvarez_02}, 
we explore the antiferromagnetic/paramagnetic transition of the  
{\em whole} family of rare-earth nickelates studied so far 
R$_2$BaNiO$_5$  R=Nd,Er,Pr 
and we propose an effective model for the behavior of these
materials in the critical region.

We have organized the paper as follows. 
 In section \ref{method} we describe the 
QMC algorithm used for our calculations and the data analysis procedure. 
In section 
\ref{critical}  we present  the critical temperatures of the 
microscopic model evaluated with QMC and we compare them 
with the prediction  given by the staggered 
mean field approach proposed by Zheludev {\it et al.} \cite{ZHELUDEV98,ZHELUDEV00}. 
An effective model for the critical behavior is presented 
 in section \ref{Effective} 
and the QMC result for the  spin-spin correlations 
in the paramagnetic phase in section  \ref{paramagnetic}.        

\section{Method and data analysis} 
\label{method} 
To study numerically the Hamiltonian (\ref{Hamiltonian}) we have used 
the Loop Algorithm \cite{EVERTZ} (see also \cite{REVLOOP} for an extensive  
review), 
a variant of the QMC method. This method belongs to the family of quantum cluster algorithms 
and provides a very efficient prescription for the 
sampling process of the configuration space. 
The key for the success of these methods is the following. 
An accurate  simulation of a spin system requires 
to gather a large sample of (nearly) statistically independent 
spin configurations in which we measure the physical 
magnitudes of interest. A simple procedure to collect a sample 
of spin configurations is by performing local updates 
involving a few spins in each Monte Carlo step. 
However, if the correlation length is sizeable, the number
of Monte Carlo steps necessary to decorrelate two spin 
configurations (also called autocorrelation time) is 
large.      
To avoid such slowing down, it is necessary to perform   
global updates involving  
clusters of spins with a size of the correlation length 
which:  a) preserve all the symmetries of the hamiltonian,
b) keep the system near equilibrium. 
The Loop algorithm gives an efficient prescription 
for constructing such clusters. In this way  
the autocorrelation time is of order one ( i.e.
a single Monte Carlo step generates a new spin configuration 
which is nearly independent from the previous one).
This is especially important since we are interested 
also in critical properties where the correlation length 
is of the size of the system.

The fact that the coupling between  S=1 and  s=1/2 is of  Ising type
and not Heisenberg simplifies 
substantially the implementation of the algorithm.    
On the other hand, the Hamiltonian (\ref{Hamiltonian}) 
is not frustrated  and does not show the
sign problem,  therefore all  Boltzmann weights appearing in the
evaluation of thermodynamic properties can be taken positive 
after the conventional rotation around the z-axis of all the spins 
in one of the two sublattices of the S=1 system. 
The temperatures of interest for comparison with experiment 
are, as we will see, well inside the scope of our QMC method.

\begin{figure}
\vspace{0.4cm}
 \includegraphics[width=0.4\textwidth]{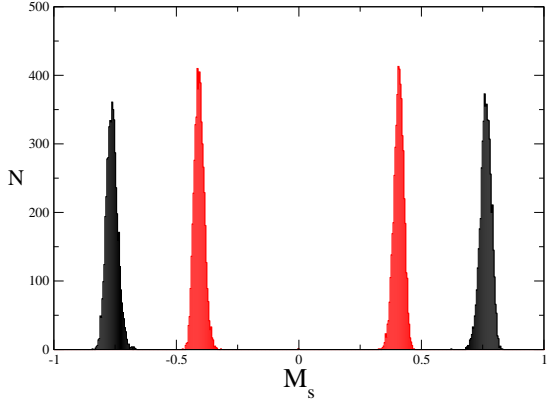}
 \caption{\label{hist_mag} Histograms of the magnetization 
  in the ordered phase for both sublattices 
   and $J_{c}=0.31$ and $T=0.1J$. The black (grey) histogram
 corresponds to the S=1 (s=1/2) species. N is the number of Monte Carlo 
 measurements of the staggered magnetization giving as a result $M_{s}$ 
 after thermalization. The figure suggests that the model 
 has two ground states.}
 \end{figure}

The
QMC simulations were performed on finite lattices and we carried out a 
 finite size scaling in order to be able to compare with experiments and
 we considered periodic
boundary conditions ${\bf S}_{L+1,2j}={\bf S}_{1,2j}$ and
${\bf S}_{i,2j}={\bf S}_{i,1}$. In order to illustrate the finite
size analysis, in Figure (\ref{hist_mag})
we show the distribution of the staggered value of the magnetization 
after 10$^5$ Monte Carlo steps.


Since the model has two ground states in the two sublattices in the ordered phase related by
the $Z_2$ symmetry
 (see Fig.(\ref{hist_mag})), 
the staggered magnetization  was 
computed  by taking the absolute value of the staggered 
component of the spin operator in each Monte Carlo step  in order to
avoid
averaging between configurations around these two ground states,  and then  
using the relation 
\begin{eqnarray*}
M_{Ni}&=&\lim_{L \rightarrow \infty }\frac{4}{L^2}\sum_{N_{MC}}|\sum_{ij}(-1)^i S_{i,2j}| \\
M_{Nd}&=&\lim_{L \rightarrow \infty }\frac{4}{L^2}\sum_{N_{MC}}|\sum_{ij}(-1)^i s_{i,2j-1}|
\end{eqnarray*}

where $N_{MC}=10^5$ is the number of Monte Carlo steps after thermalization.
In principle, we considered for this extrapolation L$\times$L lattices of size
L=8, 16 ,24 , 32 ,48, 64 spins, where L is the total number of spins including 
both magnetic species.


\begin{figure}
\vspace{0.4cm}
 \includegraphics[width=0.4\textwidth]{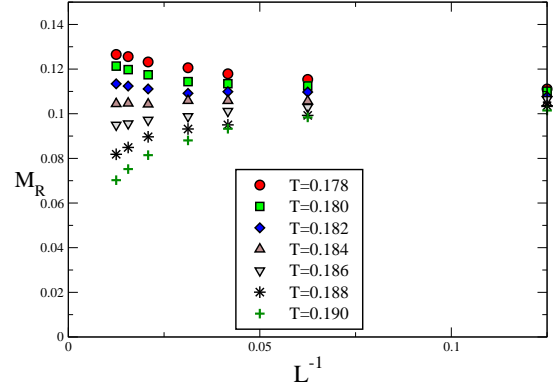}
 \caption{\label{M_L} Magnetization in the rare-earth sublattice as a 
  function of the inverse size of the system for temperatures very close 
   to the N\'eel temperature $T_N \sim 0.183$ in units of J. 
   The value of the transverse 
   coupling here is $J_{c}=0.36J$. Note that the $T=0.184$ and $T=0.186$
   seem to be in the paramagnetic phase  
   which is only observable in very large lattices (M$_R$ $\rightarrow$
0 as L $\rightarrow$ $\infty$).}
\vspace{0.6cm}
 \end{figure}

\begin{figure}
\vspace{0.4cm}
 \includegraphics[width=0.4\textwidth]{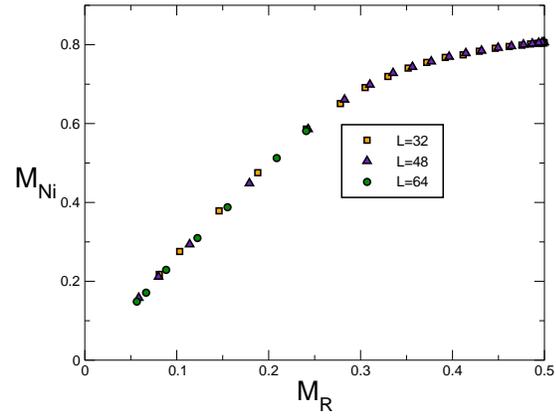}
 \caption{ \label{M_R_Ni_L} Staggered magnetizations in both sublattices 
   for different values of L . 
   Far from the critical point, where the correlation 
   length is small, the extrapolation is straightforward. 
For a given system size each point corresponds 
to a different temperature in the ordered regime
$T<T_N$.}
 \end{figure}

In Fig. (\ref{M_L}) we show the magnetization
in the rare-earth sublattice as a function of the inverse size of the
system for a model with $J_c$ = 0.36J where $T_N$ = 0.183J. 
We observe that for temperatures slightly above $T_N$, i.e. in the
paramagnetic phase, very large lattices are necessary to extrapolate 
the correct value of the magnetization (zero in this case).   
 However,  the extrapolation is an issue
only at temperatures very close to the $T_{N}$.  
As we will see, for comparison with experiments 
we need to compute the staggered 
magnetization at temperatures clearly smaller than $T_{N}$ where 
extrapolation is straightforward. Actually, that is
what we see in Figure (\ref{M_R_Ni_L}) where the relation between the
staggered magnetizations in both sublattices for different L values is shown.  In this case the 
staggered magnetization data may already have converged 
with lattices as small as 32 $\times$ 32.


\begin{figure}
\vspace{0.4cm}
 \includegraphics[width=0.4\textwidth]{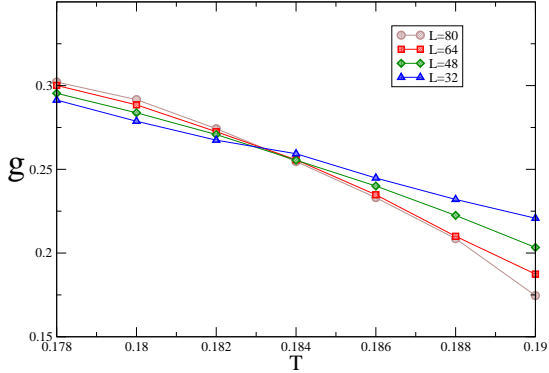}
\vspace{0.2cm}
 \caption{Binder Parameter g as a function of the temperature for 
$J_c=0.36$. The line intersection for different system sizes
signals the N\'eel temperature. \label{FSS_mag_Jc0.36} }
 \end{figure}

The N\'eel temperature was 
determined by using the Binder parameter \cite{BINDER00} $g$, which is
the fourth cumulant of the order parameter distribution. 
\be
g=\frac{3}{2}(1-\frac{\langle M^4 \rangle}{\langle M^2 \rangle}) 
\ee
At $T_N$ this cumulant  is independent of the size of the system, apart
from subdominant corrections to the critical scaling. 
In Fig.  (\ref{FSS_mag_Jc0.36}) we present the typical finite-size
scaling for the Binder parameter for $J_c = 0.36J$ computed in
lattices  
 of sizes L=24, 32, 48, 64 spins where the intersection for
different system sizes signals the N\'eel temperature. The data for L=8, 16  
show consistently significant subdominant corrections and they were not 
included in the computation of $T_N$. We suspect that this is a consequence of the complex structure of the higher e\-nergy excitations, probably     
Haldane excitations in the S=1 sector surviving in the vincinity of the 
N\'eel temperature, what sets another lengthscale $\xi_{ch} \sim 6 $ slightly 
smaller than the correlation length of an independent S=1 Heisenberg chain.
Actually, as we will see below,  the energy scale associated  
to these excitations ( the ``gap'') increases as we increase $J_c$.
Therefore only sizes $L\gg \xi_{ch}$ should enter the finite size analysis
\cite{unities}.


\section{Numerical simulations {\em vs.} mean field approach 
in the critical regime}
\label{critical}

The critical properties deserve special attention because 
the effective staggered field approach proposed by Zheludev {\it et al.}
\cite{ZHELUDEV98,ZHELUDEV00}
-being a mean field approximation-
can break down in the vincinity of the critical region, i.e. at the N\'eel temperature.
%
 The idea of the mean field is that the behavior
of the  R$_{2}$BaNiO$_{5}$ compounds in the ordered phase can be
described in terms of a $S=1$ chain of Ni ions in a staggered
magnetic field induced by the magnetic rare earth
ions: 
\be
M_{\rm Ni}={\mathcal M}(\alpha M_{\rm R} ) 
\label{magnetization_1}
\ee
where ${\mathcal M(h)}$ is the staggered magnetization of 
a S=1 spin Heisenberg chain as a function of the external staggered 
magnetic field $h$ induced by the s=1/2 ions and $\alpha$ is the
proportionality
constant. 
Inversely, the otherwise free spins $s=\frac{1}{2}$
see the mean-field produced by the neighboring $S=1$ chains.
The staggered magnetization of the R lattice is then related to 
the staggered magnetization of the Ni lattice by the expression.
\be
M_{\rm R}=M_0 \tanh (\alpha \beta M_{\rm Ni}) \\
\label{magnetization_12}
\ee
where   $M_0$ is the effective moment of the rare
earth ion and  $\beta = 1/k_B T$.

\begin{figure}
\vspace{0.4cm}
\includegraphics[width=0.4\textwidth]{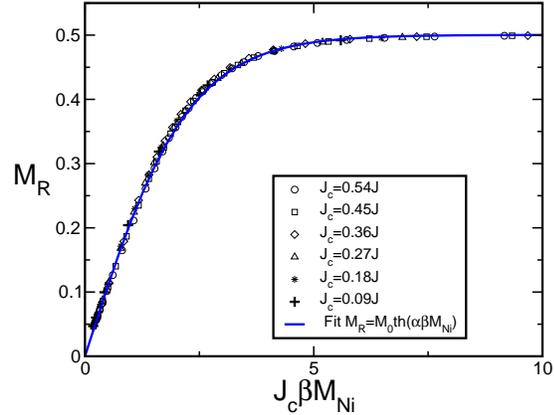}
  \caption{\label{fits} 
 Functional dependence of the QMC results for $M_{\rm R}$ vs. $M_{\rm Ni}$ for various values of
the coupling constant $J_c$. The
solid lines correspond to  Eq.\
(\protect\ref{magnetization_12}).}
\vspace{0.5cm}
\end{figure}

\begin{figure}
\vspace{0.3cm}
 \includegraphics[width=0.4\textwidth]{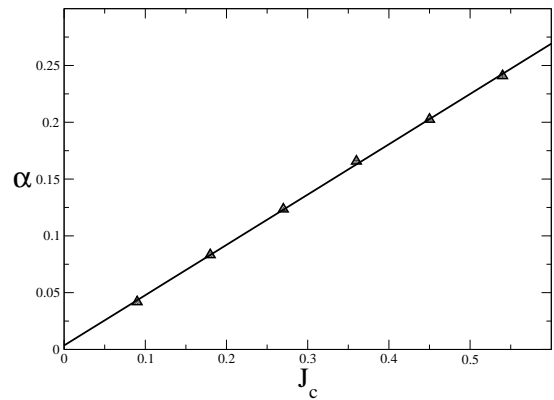}
 \caption{Effective coupling $\alpha$ between the two sublattices
as a function of the Ni-R coupling as obtained from the fits
in Fig. \ref{fits} The solid line is a fit to the linear law 
$\alpha = 0.443J_c+0.0034$.  \label{alpha_Jz}}
 \end{figure}

We aim now to compare our QMC numerical results with the predictions for
$T_N$ of the staggered mean field approach.  
The coupling between the two sublattices in the staggered mean 
field formalism is encoded in the constant $\alpha$ instead of the 
microscopic coupling constant $J_c$. Therefore the first step is to find
a relation between both in the antiferromagnetic phase.    
In Fig.\ (\ref{fits}) we present the  QMC results for  $M_{R}$ =  $f(J_c
\beta M_{Ni})$   for various values
of $J_c$.  Note that all curves fall on top of each other and, in fact,
this
curve can be described  by Eq. (\ref{magnetization_12}) where 
 $\alpha$ must be  a linear function of $J_c$ in the range
of transverse couplings studied as observed from our QMC results.
Furthermore,  we have extracted the relation $\alpha=0.0034+0.443J_c$
in that range, which can be considered linear 
within our statistical error bars (see Figure \ref{alpha_Jz}). 

\begin{figure}
\vspace{0.5cm}
\includegraphics[width=0.4\textwidth]{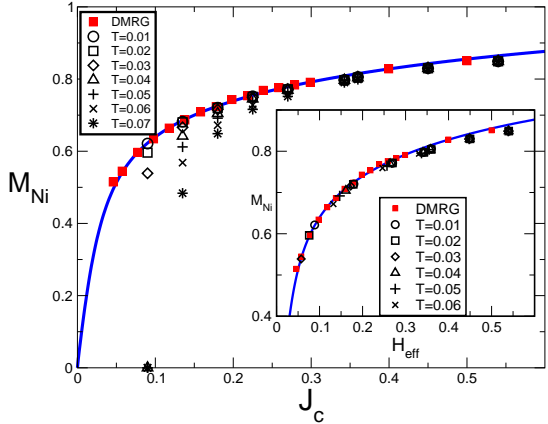}
\caption{\label{M-H} $M_{\rm Ni}$ vs. the value of 
the $J_{c}$ for several values of $J_c$ and  temperatures in units of $J$.
In the inset $M_{\rm Ni}$ vs. the
'effective' field on the Ni-subsystem induced by $M_{\rm R}$
$H=2J_cM_{\rm R}$ \protect\cite{ZHELUDEV98} for various T values. The squares
correspond to the DMRG results \protect\cite{YU99} at T=0.}
\end{figure}

At this point it is legitimate to question the relation between 
a microscopic coupling $J_c$ and a frankly phenomenological 
constant $\alpha$ {\em especially at finite temperatures}.
Since our QMC data are collected in the range of temperatures
$T_{\rm min}<T<T_N$, where $T_{\rm min}$ was the minimal energy that 
we were able to simulate,  it is necessary to study whether there 
is a temperature dependence in $\alpha$.  
In the main picture of Fig.\ (\ref{M-H}) we show the relation
between $M_{\rm Ni}$ and the value of the transversal coupling 
$J_c$ at various temperatures.   Actually the data converge in the T=0 limit to a well
defined curve $M_{\rm Ni}(J_c)$ which at low values of $J_c$
is the magnetization curve of the 1D S=1 Heisenberg model 
at T=0 in a staggered magnetic field computed by Yu {\it et al.}
\cite{YU99} using 
the Density Matrix Renormalization Group (DMRG) method.

In the inset of Fig.\ (\ref{M-H}) we directly plot $M_{\rm Ni}$ vs.
$H_{\text{eff}}$ for various T values assuming that $\alpha$ is temperature
independent. Now we have that even when  
the staggered magnetization in the R lattice is not at saturation   
we find that the rescaled data are superimposed to 
the magnetization curve of the 1D S=1 Heisenberg model at low values of $J_c$.
This data collapse is a consequence  
of the large value of the Haldane gap 
and the low temperatures considered. At higher values of $J_c$, 
the R sublattice remains nearly saturated at the highest values of the T 
considered (T=0.07J), therefore no difference is observed when the 
rescaling is done. The results presented in Fig.\ (\ref{M-H}) 
also confirm that the assumption of a temperature independent $\alpha$
is valid in the range of temperatures considered and 
that the mean field approach works very well  
if the chains are weakly coupled to the R magnetic moments 
but there are small deviations at higher values of $J_c$. Analyzing
the QMC results,  the scale that separates the 1D 
from the 2D regime is the Haldane gap 
$H_{\text{eff}} \sim \Delta_{1D} \sim 0.41J$ as expected. We also note 
that  for the smallest value of $J_c$,  $J_c=0.09J$ 
the N\'eel temperature is so small $T_N=0.06J$  
that some of the points are still in the 
paramagnetic phase $M_{\rm Ni}(T)=0$. 
      
\begin{figure}
 \includegraphics[width=0.38\textwidth]{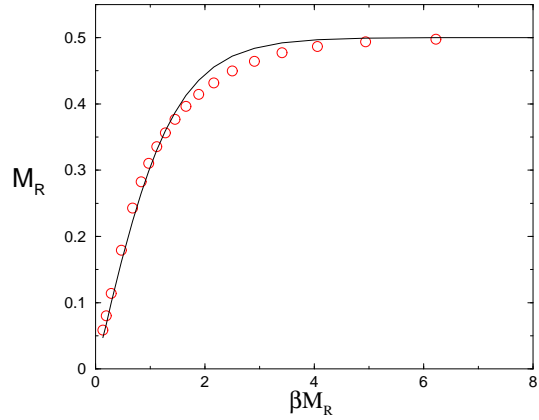}
\vspace{0.2cm}
 \caption{\label{No_MF}
{\em Classical} mean-field (solid line) compared to QMC results
  (open circles). Note that a classical mean field 
theory approach fails to correctly relate 
the staggered magnetization in any sublattice. Here we show the best 
possible fit using the rare-earth subsystem.}
\vspace{0.5cm}
 \end{figure}

In addition, to corroborate the relevance of the quantum fluctuations 
in this system, we present in Fig. (\ref{No_MF}) a comparison
 of a classical mean field model  
of the staggered magnetization in the R
sublattice  
i.e. $M_{\rm R}$ as a function of $\beta M_{\rm R}$ and independent
of the magnetization of the Ni sublattice (solid line) (Brillouin function) with   the QMC data
obtained with the Hamiltonian (\ref{Hamiltonian}) (open circles).  We observe
that there is a significant deviation between the data and the 
Brillouin function which can be attributed 
to the effect of the quantum fluctuations on the magnetization in
the R sublattice which are contained in the model Eq. (\ref{Hamiltonian})
but not in a classical mean field model.   


\begin{figure}
\vspace{0.5cm}
 \includegraphics[width=0.4\textwidth]{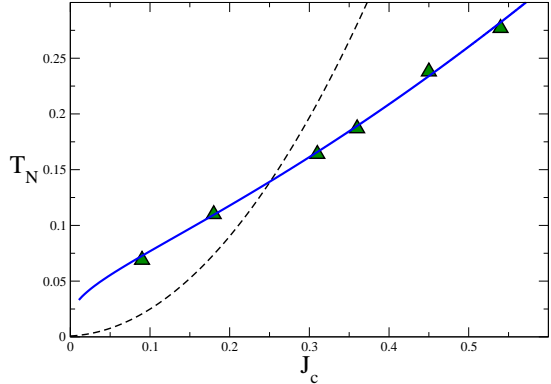}
 \caption{ N\'eel temperature as a function of the Ni-R coupling 
$J_{c}$ in units of $J$. The triangles are the result of our QMC simulation in our 
model Hamiltonian. The dashed line is the 
mean field result given by the equations 
(\ref{magnetization_12}-\ref{magnetization_1})
and the solid line is the fit to a 2D anisotropic 
Ising model \label{T_N-J_c} (see text).}
 \end{figure}

A numerical solution of Eqs. (\ref{magnetization_1}-\ref{magnetization_12})
in the whole AFM phase ( $0 \leq M_{\rm R} \leq 0.5$ ,
$0 \leq M_{\rm Ni} \leq 1.0$ ) demands an explicit analytic expression
for the staggered magnetization as a function of the 
external staggered magnetic field in a S=1 chain.  
For that purpose 
we took the DMRG results of Ref. \cite{YU99} and got  
the best possible fit.
  The only  conditions that we impose to 
our fitting function are: i) It should have a 
smooth fitting behavior in the wide range of
values necessary to the numerical solution of
(\ref{magnetization_1}-\ref{magnetization_12})
ii) it should show linear behavior at low external fields and  iii) 
it should have asymptotic saturation at high  external fields.
The explicit expression that we obtained was:
   \be
      M=A\arctan(Bh)+C\tanh(Dh)+Eh/(1+Eh);
   \ee    
where A=0.177457, B=9.58055, C=0.336168, D=56.5282 and E=0.386596. 
The equations  (\ref{magnetization_1}-\ref{magnetization_12})
can be solved now numerically  confirming in general the excellent 
agreement of QMC and the mean field in the ordered phase but  
there is {\it qualitative} disagreement in the $T_N(J_c)$ relations
 as shown in Fig. (\ref{T_N-J_c}).  In  Fig. (\ref{T_N-J_c})
we compare the N\'eel temperature as a function of $J_c$ 
obtained numerically (triangles) using the Binder parameter as explained in the
previous
section with the mean field solution (dashed line). 
The most significant feature of 
the mean field solution is its  quadratic behavior 
in $\alpha$ and therefore in $J_c$. This result suggests
that the functional relation between $\alpha$ and $J_c$ changes 
abruptly close to the critical point and $\alpha \sim \sqrt{J_c}$ 
would be more appropiate in that regime. We shall investigate this
discrepancy
in more detail in the next section.


 \section{Effective description of the critical behavior}     
\label{Effective}

Our purpose here is to
 reproduce 
the relation $T_N(J_{c})$  (see Fig. (\ref{T_N-J_c})) using a model that includes the 
critical fluctuations that seem to be missing 
in the mean field approach.

 To elaborate such a model let us consider the following observations: 

i.) The magnetic moments in the rare earth conmute with 
the total Hamiltonian
\be
[s^{z}_{i, 2j\pm 1},H]=0 
\ee
which implies that $M_{\rm R}$ is a good quantum number of the 
Hamiltonian irrespective of the value of $J_{c}$.  

ii.) At finite temperatures the model has two ground states related by the 
$Z_2$ symmetry in both sublattices 
as shown in the histograms of the magnetization in the ordered region (see
Figure (\ref{hist_mag}) where the distribution of the order parameter 
is presented after thermalization.) 

iii.) The spin rotational invariance has been broken by the Ising 
character of $J_c$. As predicted by Haldane \cite{HALDANE83} 
and elaborated on microscopical grounds by Gomez-Santos
\cite{GOMEZ-SANTOS},  
the low-energy sector of the XXZ S=1 Heisenberg chain can be mapped 
onto a 2D Ising model. Actually, the longitudinal  
spin-spin  correlation function  of a Haldane chain is identical 
to the one of the 2D Ising model in the 
paramagnetic phase, including the power-law correction to the scaling.
The previous relation is an example of mapping a quantum model in dimension d
to a classical model in d+z dimensions, where z=1 for this case. The length 
of the imaginary time dimension is determined by the inverse 
temperature $\beta$.

Let us consider then a system of Ising
planes coupled by the intercalated moments of the rare-earth.
In the materials of interest we have always $T_N < \Delta$.
In the para\-mag\-netic phase  $T_N < T <\Delta$ we expect 
that the autocorrelation time $\xi_T$ 
in the imaginary direction to be smaller
than $\beta$ and the system behaves effectively as a three dimensional 
Ising model. As the temperature is reduced,   a crossover takes place 
when $\xi_T$ becomes of the order of $\beta$. Beyond that point    
the system belongs to the 2D Ising model universality class  with different 
coupling constants in the  two spatial dimensions. 
    
The simplest effective model ( i.e. valid in the vincinity of the critical 
point) and  compatible with the arguments above  
is an anisotropic 2D Ising model whose 
critical temperature is given by the  Kramers-Wannier 
expression \cite{KW_41}:
\be   
     \sinh(2J_{\parallel}/T_{N})\sinh(2J_{\perp}/T_{N})=1 
\ee 
where $J_{\parallel}=2.35$ and $J_{\perp}=0.022J_c^2$ are the effective
couplings 
of the model. In Fig. \ref{T_N-J_c} we show the agreement of this
effective
model (solid line) with the QMC data obtained for Eq. (\ref{Hamiltonian}) (triangles).

\section{The paramagnetic phase} 
\label{paramagnetic}

For completion, 
 we present in this section the spatial spin-spin correlations in the
paramagnetic phase. 
At $T=T_N$ the spin-spin correlation function decays algebraically 
and therefore the correlation length is infinite. As the temperature
increases,  
the correlation length  decreases. If the nominal gap $\Delta$ of the 
independent S=1 is large enough, the properties of the independent S=1 
should become visible in the paramagnetic phase. The question we are interested in is whether this crossover is observable in the value of the correlation length.

\begin{figure}
\vspace{0.5cm}
 \includegraphics[width=0.4\textwidth]{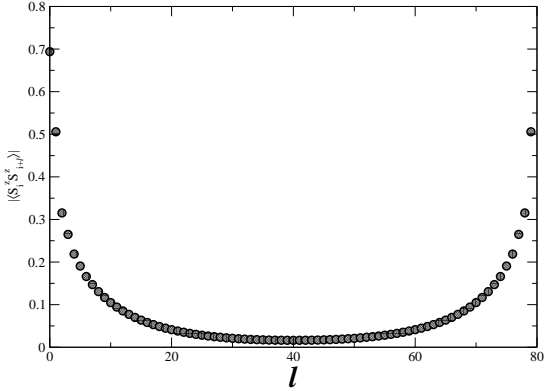}
\vspace{0.3cm}
 \caption{\label{SzSz_Jz0.18} Spin-spin correlation function (z-component)
in the Ni subsystem and along the chains ($J_c=0.18$) in a lattice of 80x80 sites}
\vspace{0.5cm}
 \end{figure}
\begin{figure}
 \includegraphics[width=0.4\textwidth]{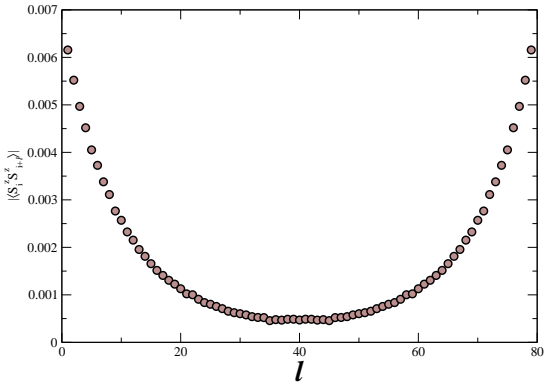}
\vspace{0.3cm}
 \caption{\label{R_SzSz_Jz0.18} Same magnitudes and parameters as 
  in figure (\ref{SzSz_Jz0.18}) for the R subsystem. Note the change in scale.  }
 \end{figure}

In Figures (\ref{SzSz_Jz0.18}) and  (\ref{R_SzSz_Jz0.18}) we present 
the correlation functions of the z-component of the spin operators 
in the Ni and R sublattice respectively. We observe that, eventhough 
the  R-R correlations are two orders of
 magnitude smaller than the Ni-Ni correlations,  there are finite correlations  between the 
s=1/2 spins mediated by the chains. We recall that the model doesn't 
couple the s=1/2 spins directly.  
The inmediate question is: which is the induced correlation length between the 
rare earth magnetic moments?

Following White and Huse \cite{WHITE93}, we calculated the correlation
length  $\xi$ for both sublattices in a  64$\times$64 spins lattice by fitting the z-component of the spin-spin correlation function to
the law:
\begin{eqnarray}
|\langle S^{z}_0 S^{z}_l \rangle| =
       A\exp(\frac{-l}{\xi})l^{-\eta}
\label{xi}
\end{eqnarray}
where $A$ and $\eta$ are fitting parameters. 
In the vicinity but not too close to $T_N$ we find  a fit 
of the correlation length temperature to a power law in 
both sublattices. We took values  in a range  $\xi \sim 5-17$ that can be 
accurately computed with the procedure described above. The best fit 
 is 
$\xi^{-1} =K(T-T_N)^{\frac{1}{2}}$ where 
 K is the only free parameter\cite{Alvarez_02}.
Such behavior is the one expected for a Ginzburg-Landau like description
{\em above} the critical point  
confirming that a mean field description is again valid as we depart 
from the critical point from above.  
Besides, we note that for $T >> T_N$, the correlation
length in the Ni subsystem approaches the correlation length of a
single chain.

 The fitting parameter $\eta$ shows a differenciated temperature 
dependence in both
sublattices i.e   $\eta=0$ for the correlation function Eq. (\ref{xi})
in the R-sublattice while in the Ni sublattice
$\eta$ approaches the value  $0.5$ as the temperature is reduced
in the paramagnetic phase.


\section{Conclusions.}
\label{conclusions.}
 
In conclusion, we have provided a detailed analysis of 
 the antiferro-/paramagnetic transition
 of the  R$_2$BaNiO$_5$ mixed-spin quantum antiferromagnets based on  
 QMC si\-mulations of a lattice model of interacting spin-1
 and spin-1/2 entities. We give an extensive description
of the numerical calculations and data analysis and comparison
 of our results with experimental observations is very good.
These calculations go beyond previous mean
 field type of approaches and retain the main features of the
 behavior of the system in the
 critical region.  
We  propose an effective 2D anisotropic Ising model
to explain the behavior of the N\'eel temperature 
as a function of the coupling between the nickel 
and the rare earth sublattices. 

\section{Aknowledgements.}
It is a pleasure for us to acknowledge discussions with C.
Gros, S. Moukouri, H. Rieger and A. Zheludev.




\begin{thebibliography}{99}
\bibitem{Alvarez_04} see J.V. Alvarez, H. Rieger and A. Zheludev,
        Phys. Rev. Lett. {\bf 93}, 156401 (2004)
        and references therein

\bibitem{Lewenstein_04} see M. Lewenstein, L. Santos, M. A. Baranov, and
                        H. Fehrmann, Phys. Rev. Lett. {\bf 92}, 050401 (2004).

\bibitem{ZHELUDEV98} A.~Zheludev, E.~Ressouche,S.~Maslov T. ~Yokoo
S.~Raymond J.~Akimitsu
 Phys. Rev. Lett. {\bf 80},  3630  (1998).

\bibitem{Alvarez_02} J.V. Alvarez, R. Valent\'\i, and A. Zheludev,
                     Phys. Rev. B, {\bf 65}, 184417 (2002).

\bibitem{HALDANE83} F.D.M. Haldane, Phys. Rev. Lett. {\bf 50},  1153
                     (1983).

\bibitem{RAYMOND99} S. Raymond, T. Yokoo, A. Zheludev, S. E. Nagler, A. Wildes and J. Akimitsu,  Phys. Rev. Lett. {\bf 82},
                    2382 (1999).









\bibitem{ZHELUDEV00} A. Zheludev, S. Maslov, T. Yokoo, S. Raymond, S. E.
	Nagler and J. Akimitsu J. Phys.: Condens. Matter {\bf 13},  R525
	(2001).

\bibitem{Maslov_98} S. Maslov and A. Zheludev, Phys. Rev. Lett. {\bf
	           80}, 5786 (1998).


\bibitem{YBaNiO}  J. Darriet and L. P. Regnault, Solid State Commun. {\bf 86}, 409 (1993);T. Yokoo, T. Sakaguchi, K. Kakurai and J. Akimitsu, J. Phys. Soc. Japan {\bf
  64}, 3651, (1995); G. Xu, J. F. DiTusa, T. Ito, K. Oka, H. Takagi, C. Broholm, G. Aeppli,  Phys. Rev. B {\bf 54}, R6827 (1996).



\bibitem{ALONSO90}J.A. Alonso, J. Amador, J.L. Martinez, I. Rasines ,J.  Rodriguez-Carvajal,R. Saez-Puche, Solid State Commun. {\bf 76}, 467 (1990).

\bibitem{GARCIA-MATRES93}  E. Garc\'{\i}a-Matres, J.L. Mart\'{\i}nez,
          J. Rodr\'{\i}guez-Carvajal, A. Salinas-S\'anchez, Solid State
Commun.
          {\bf 85} , 553 (1993).


\bibitem{ZHELUDEV96} A. Zheludev, J. M. Tranquada, T. Vogt, D. J. Buttrey, 
                    Phys. Rev. B. {\bf 54},7210 (1996).




\bibitem{EVERTZ} H.G. Evertz, G. Lana and M. Marcu,
                 Phys. Rev. Lett. {\bf 70}, 875 (1993)

\bibitem{REVLOOP} H.G. Evertz, Adv.Phys. {\bf 52},  1 (2003)





\bibitem{BINDER00} {\it Monte Carlo Simulations in Statistical Physics}
                   D. P.
                 Landau and K. Binder (Cambridge University Press,
        2000).
\bibitem{unities} Calculations are done in units $\mu_{B}=k_{B}=\hbar=1$.



\bibitem{YU99} ~J.~Lou, ~X.~Dai,~S.~Qin,~Z.~Su,~L.~Yu.
              Phys. Rev. B {\bf 60}, 52 (1999).

\bibitem{GOMEZ-SANTOS} G. Gomez-Santos, Phys. Rev. Lett. {\bf 63}, 790
	(1989)

\bibitem{KW_41} H.A. Kramers and G.H. Wannier, Phys. Rev. {\bf 60}, 252  (1941).

\bibitem{WHITE93} S. R. White and D. A. Huse, Phys. Rev. B {\bf 48}, 3844
                  (1993).




%
%

%
%
%


\end{thebibliography}
\end{document}